\documentclass[aps,prl,showpacs,twocolumn]{revtex4}

\usepackage{graphicx}
\usepackage[ansinew]{inputenc}
\usepackage{color}
\usepackage{float}
\definecolor{Blue}{rgb}{0.3,0.3,0.9}
\begin{document}
\topmargin 0.6cm

\title{Photoemission footprints of extrinsic plasmarons}
\author{B. Hellsing$^{1,2}$}
\email{hellsing@physics.gu.se}
\author{V.M. Silkin$^{2,3,4}$}

\affiliation{$^1$Department of Physics, Gothenburg University, S-41296 Gothenburg, Sweden \\
$^2$Donostia International Physics Center (DIPC), 20018 San Sebasti\'{a}n, Spain\\
$^3$Depto. de F\'{\i}sica de Materiales, Facultad de Ciencias Qu\'{\i}micas, Universidad del Pa\'{\i}s Vasco, Apdo. 1072, 20080 San Sebasti\'an,  Spain\\
$^4$IKERBASQUE, Basque Foundation for Science, 48011 Bilbao, Spain }

\begin{abstract}
A prediction how to experimentally distinguish excitations of extrinsic plasmarons from intrinsic plasmarons is presented. In surface systems where excitations of acoustic surface plasmons is possible it is shown that the photo-electron yield in normal photoemission should decay according to an inverse square root dependence with respect to the photon energy. A computational analysis of the system p(2$\times$2)-K/Graphite confirms this prediction.
\end{abstract}


\maketitle

\section{Introduction}

In photoemission experiments, photo-electrons carry information of many-body interactions created by the photo-hole and by the escaping photo-electron itself. In the case of strong coupling between the photo-hole or induced density caused by the photo-electron, and plasmon excitations, the quasi particle picture breaks down and new loss peaks appear in the photoemission spectrum. The excitation formed by the photo-hole -- plasmon interaction defines the $intrinsic$ plasmaron and the photo-electron -- plasmon interaction defines the $extrinsic$ plasmaron \cite{Lundqvist_PKM67,helussc67}.

For systems with a surface-state band crossing the Fermi level, a plasmon localized at the surface and characterized by a sound-like dispersion, so-called Acoustic Surface Plasmon (ASP), has been predicted to exist \cite{vsi_04,vsi_10}. Later on Electron Energy Loss Spectroscopy experiments have confirmed the presence of the ASP mode at the Be(0001) \cite{bdi_07,jamuprb12} and noble metal surfaces \cite{spa_10,podiepl10,vavep12,yajaprb12,vasmprl13,piwejpcc13,smvaprl14, pochpss15}  in good agreement with calculations and graphene adsorbed on metal substrates \cite{liwiprb08,pflajpcm11,shhwprb11,lafonjp11,lapfnjp12,pomaprb11,pomaprb12,cuposs15}.
In cases when surface localized quantum well states are formed, e.g. when atomic layers of alkali metals are adsorbed on a metal surface the possibility opens up to design ASP by varying the depth of the quantum well (type of alkali atoms) and the width of the quantum well (number of layers).

A challenge is to find out about the relative occurrence of the {\it intrinsic} and {\it extrinsic} plasmarons from a photoemission experiment. In this paper we show that for a surface system with ASP, the extrinsic plasmaron excitation channel can be traced by looking at the photon energy dependence of the photo-electron yield. Simple kinematics indicate an inverse square root dependence, while the {\it intrinsic} plasmaron is not expected to depend on the photon energy. We will present results from an extension of a previous calculation on the system - a monolayer potassium adsorbed on graphite (p(2$\times$2)-K/Graphite) \cite{Vasse}.

\section{Theory}

The intention is to form a theory in order to calculate the yield of extrinsic plasmaron excitations. The energy loss induced by the escaping photo-electron in an ARPES experiment is given by the rate of electronic surface excitations. In first order time dependent perturbation theory, "golden-rule", we have that the rate of $extrinsic$ plasmaron excitation is given by \cite{Vasse}

\begin{eqnarray}
\label{eq:GR4}
W(\omega) = -2 \ {\rm Im}  \left[ \int {\rm d}{\bf r} \phi_{ext}^{*}({\bf r},\omega) \delta n({\bf r},\omega)\right],
\end{eqnarray}
%
%
where $\phi_{ext}$ is the external potential created by the escaping photo-electron and $\delta n$ is the induced electron density.

We consider an ARPES experiment and the possibility that an ejected photo-electron will lose part of its energy on its way to the detector. We assume that the electron density "spill-out" from the surface is neglectible  along the path of the photo-electron. The external potential must then fulfill Laplace equation \cite{Liebsch_97}
\begin{eqnarray}\
\label{eq:ext}
\phi_{ext}({\bf r}_{\|},z,\omega) =
-\frac{1}{A}\sum_{{\bf q}_{\|}}\frac{2\pi}{q_{\|}}\,e^{q_{\|}(z-\tilde{z})}e^{{\rm i}({\bf q}_{\|}\cdot{\bf r}_{\|}-\omega t)} ,
\end{eqnarray}
where $A$ is the surface area, $q_{\|}=|{\bf q}_{\|}|$ and $\tilde{z}$ is the distance between the surface and the photo-electron. The rate of energy loss can then be expressed in terms of the surface response function \( g({\bf q}_{\|},\omega) \) \cite{Persson_PRB_1984}
\begin{eqnarray}\
\label{eq:g} g({\bf q}_{\|},\omega) =
\int {\rm d}z \ e^{q_{\|}z} \delta n(z,{\bf q}_{\|},\omega)
\label{eq_g}
\end{eqnarray}
and, accordingly,
\begin{eqnarray}\
\label{eq:wfin}
W(\omega,\tilde{z}) =
 \frac{4\pi}{A} \sum_{{\bf q}_{\|}}\frac{e^{-q_{\|}\tilde{z}}}{q_{\|}} \,{\rm Im}[g({\bf q_{\|}},\omega)].
\end{eqnarray}
We consider at this point a general system with an ultra thin metal adlayer adsorbed on metal surface. We assume formation of a surface quantum well (QW) hosting a QW-state band and as a result the existence of ASP.

A photo-excited electron with an initial parallel wave vector ${\bf k}'_{\|}$ will with some probability be inelastically scattered to ${\bf k}_{\|}$ while exciting an ASP with momentum ${\bf q}_{\|}={\bf k}_{\|}-{\bf k}'_{\|}$. The photo-electrons with momentum ${\bf k}_{\|}$=${\bf k}'_{\|}$ will yield the main peak, corresponding to the electrons having absorbed fully the photon energy. The width of this elastic peak reflects the finite lifetime of the photo-hole left behind. In addition a satellite structure might appear at higher binding energies due to scattering from all ${\bf k}'_{\|}$ and ${\bf k}_{\|}$, satisfying  ${\bf k}_{\|}={\bf k}'_{\|}+{\bf q}_{\|}$, having excited an ASP with momentum ${\bf q}_{\|}$. If this satellite structure gives rise to a distinct peak a $extrinsic$ plasmaron excitation is realized.

We calculate the ${\bf k}_{\|}$-resolved photo-electron energy loss per time unit due to the ASP excitations, which is equivalent to the dispersion of the $extrinsic$ plasmaron excitations. This can be carried out from the expression given in Eq. (\ref{eq:wfin})
\begin{eqnarray}
W(k_{\|},\epsilon,\tilde{z}) &=& \frac{2}{\pi}\int_{0}^{k_{F}}{\rm d}k'_{\|} \int_{0}^{2\pi}{\rm d}\alpha \  \frac{k'_{\|}}{q_{\|}} \ e^{-q_{\|}\tilde{z}} \Theta(q_{\rm max}-q_{\|}) \   \nonumber \\
&\times& {\rm Im}[g(q_{\|},\epsilon-\epsilon_{b}+\hbar\omega(k'_{\|}))]   \ ,
\label{eq_Wkomega}
\end{eqnarray}
where  $k_{F}$ is the band Fermi wave vector, $k_{\|}=|{\bf k}_{\|}|$, $k'_{\|}=|{\bf k}'_{\|}|$, $\alpha$ the angle between the vectors ${\bf k}_{\|}$ and ${\bf k}'_{\|}$. $\Theta(x)$ is the Heaviside step function, $\epsilon_{b}$ the binding energy, $q_{\|}  = (k_{\|}^{2} + k_{\|}^{'2} - 2k_{\|} k'_{\|} cos \alpha)^{1/2}$, $q_{max}$ the maximum wave vector up to which the ASP dispersion is well defined and $\hbar\omega(k_{\|})$-$\epsilon_{b}$ is the band dispersion relative the Fermi energy.

The ARPES intensity is given by integrating in time the excitation rate $W$
\begin{eqnarray}
I(k_{\|},\epsilon; h \nu) &=& \int_{0}^{\infty} W(k_{\|},\epsilon, \tilde{z}(t))dt  \ ,
\label{eq_Wkomega}
\end{eqnarray}
but as $d\tilde{z}=k_{\perp} dt$, where $k_{\perp}$ is the perpendicular mometum of the escaping photo-electron, we can now integrate with respect to $\tilde{z}$, obtaining the photon energy dependence of the $extrinsic$ plasmaron dispersion. Simple kinematics in terms of photon energy $h\nu$, workfunction $\phi$ and QW binding energy $\epsilon_{b}$ yields
\begin{eqnarray}
k_{\perp}^{2} &=& 2( h\nu - \phi - \epsilon_{b} - \epsilon + \hbar\omega(k'_{\parallel})) - k_{\parallel}^{2} \ .
\label{eq_kperp}
\end{eqnarray}
We then have the photon energy dependent intensity of $extrisic$ plasmaron
\begin{eqnarray}
I(k_{\|},\epsilon; h\nu) &=& \frac{2}{\pi}\int_{0}^{k_{F}}{\rm d}k'_{\|} \int_{0}^{2\pi}{\rm d}\alpha \  \frac{k'_{\|}}{k_{\perp}q_{\|}^{2}} \Theta(q_{\rm max}-q_{\|}) \   \nonumber \\
&\times& {\rm Im}[g(q_{\|},\epsilon-\epsilon_{b}+\hbar\omega(k'_{\|}))]   \ .
\label{eq_I_photon}
\end{eqnarray}
The expression in Eq. (\ref{eq_I_photon}) forms the footprint of $extrinsic$ plasmaron excitations. In the case of normal photoemission (${\bf k}_{\|}$=0) when the photon energy $h\nu$ exceeds $\phi - \epsilon_{b}$ the intensity of the $extrinsic$ plasmaron excitations will decrease with photon energy accordingly, $I\sim1/\sqrt{h\nu}$. This result can be traced back to the exponential decay with respect to $\bar{z}$ (the time dependent location of the escaping photo-electron) of the external potential penetrating the solid (Eq. (\ref{eq:ext})). In the next section we illustrate this for a specific system.
\section{Calculations}

\begin{figure}
\includegraphics[width=85 mm]{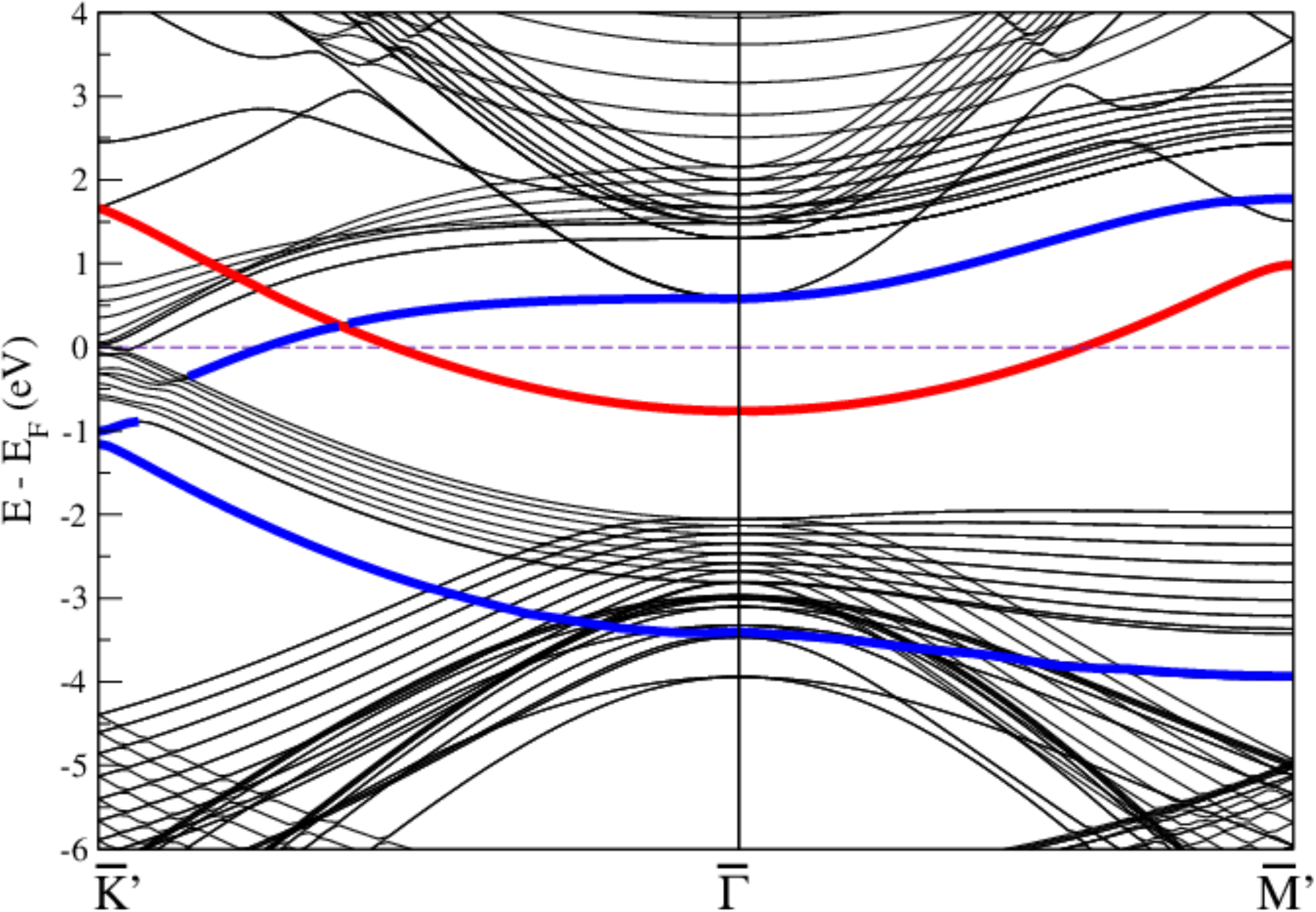}
\caption{(Color online) Calculated p(2$\times$2)-K/Graphite band structure \cite{Chis_PRB_2011}. Red and blue color lines indicate the quantum well band and the lowest and the highest branches of the folded $\pi^{*}$ and $\pi$ bands, respectively. $\bar{\text{K}'}$ and $\bar{\text{M}'}$, represent the $\bar{\text{K}}$ and $\bar{\text{M}}$ points of the folded band structure due to the (2$\times$2) overlayer of potassium.}
\label{fig_bands}
\end{figure}

With this theoretical background we proceed to a specific system, a monolayer of potassium
on graphite, p(2$\times$ 2)-K/Graphite. According to the first principles calculations by
Chis {\it et al.} \cite{Chis_PRB_2011} a quasi-2D quantum well (QW) system is formed with an energy band centered at the $\bar{\Gamma}$-point of the Brillouin zone (BZ) (see red colored line in Fig. \ref{fig_bands}). Another quasi-2D system is formed in the carbon atomic layer below the QW system, marked by blue color. Due to the larger Fermi surface of the QW band this band will contribute the most to the excitations. Thus our analysis is focused on the QW band. It should be noted however, that in the calculation of the surface response function \( g({\bf q}_{\|},\omega) \) we include all excitations.
%
%
\subsection{Acoustic surface plasmons}
\begin{figure}
\includegraphics[angle=0,origin=c,width=80 mm]{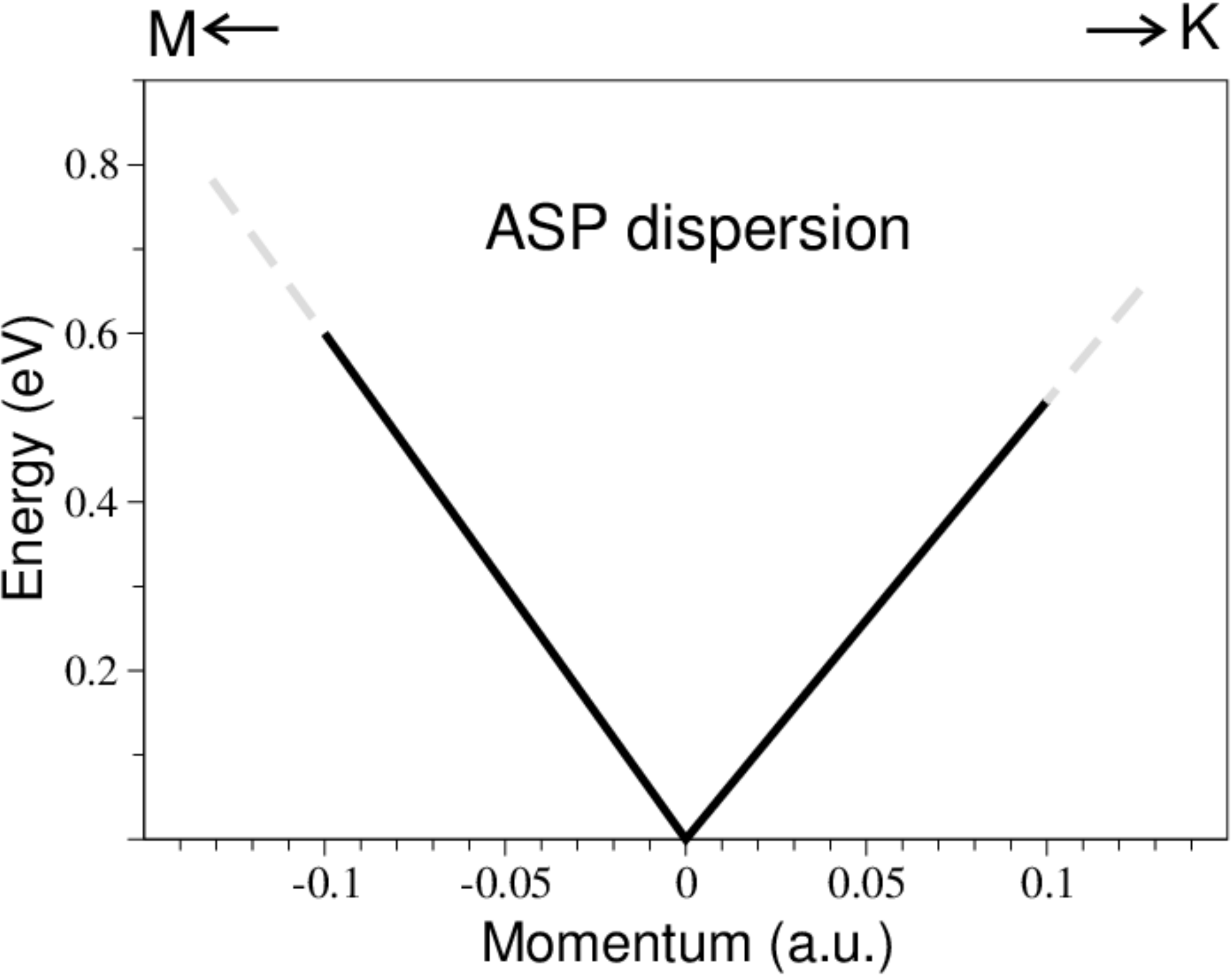}
\caption{ASP dispersion of the system p(2$\times$ 2)-K/Graphite. The arrows indicate the directions $\bar{\Gamma}$-$\bar{M}$ and $\bar{\Gamma}$-$\bar{K}$. The solid lines are extracted from the calculated Im [\( g({\bf q}_{\|},\omega) \)] in \cite{Vasse}. The dashed lines indicate the strong damping of the ASP due to incoherent excitations of electron-hole pairs.}
\label{fig_ASP_disp}
\end{figure}

We have previously calculated the surface loss function, Im [\( g({\bf q}_{\|},\omega) \)] within the Time Dependent Density Functional Theory scheme \cite{Vasse}. The surface loss function versus $\omega$ and $\bf q$ reveals a linear sound-like dispersion $\omega(\bf q)$ indicating the existence of ASP as a well-defined collective excitation in the energy range 0-0.6 eV with a momentum transfer span up to about 0.1 a.u. The extracted dispersion is shown Fig. \ref{fig_ASP_disp}. At larger momentum transfers, where the ASP dispersion is depicted by the dashed lines, this mode becomes strongly damped. Beyond this region for $q_{\|}$ $> \sim$0.13 a.u. it ceases to exist since
the coherence of single electron excitations forming the collective plasmon excitation is lost due to incoherent electron-hole pair excitations. In Fig. \ref{fig_ASP_disp} one can notice that the ASP dispersions along the $\bar{\Gamma}$-$\bar{\text{M}}$ and $\bar{\Gamma}$-$\bar{\text{K}}$ directions are very similar. Based on this observation we will further on assume that the ASP dispersion is isotropic in the surface plane.

The average slope of the ASP dispersion yields a group velocity of $c$ $\approx$ 0.22 a.u., which according to Pitarke {\it et al.} \cite{pinaprb04}, is set by the Fermi velocity of the 2D carriers. This is consistent with the band structure in Fig. \ref{fig_bands} where the slope of the QW band when crossing the Fermi level is similar, $v_{F}$ $\approx$ 0.23 a.u. \cite{Vasse}.

\subsection{Plasmaron excitations}
\begin{figure}
\includegraphics[angle=0,origin=c,width=80 mm]{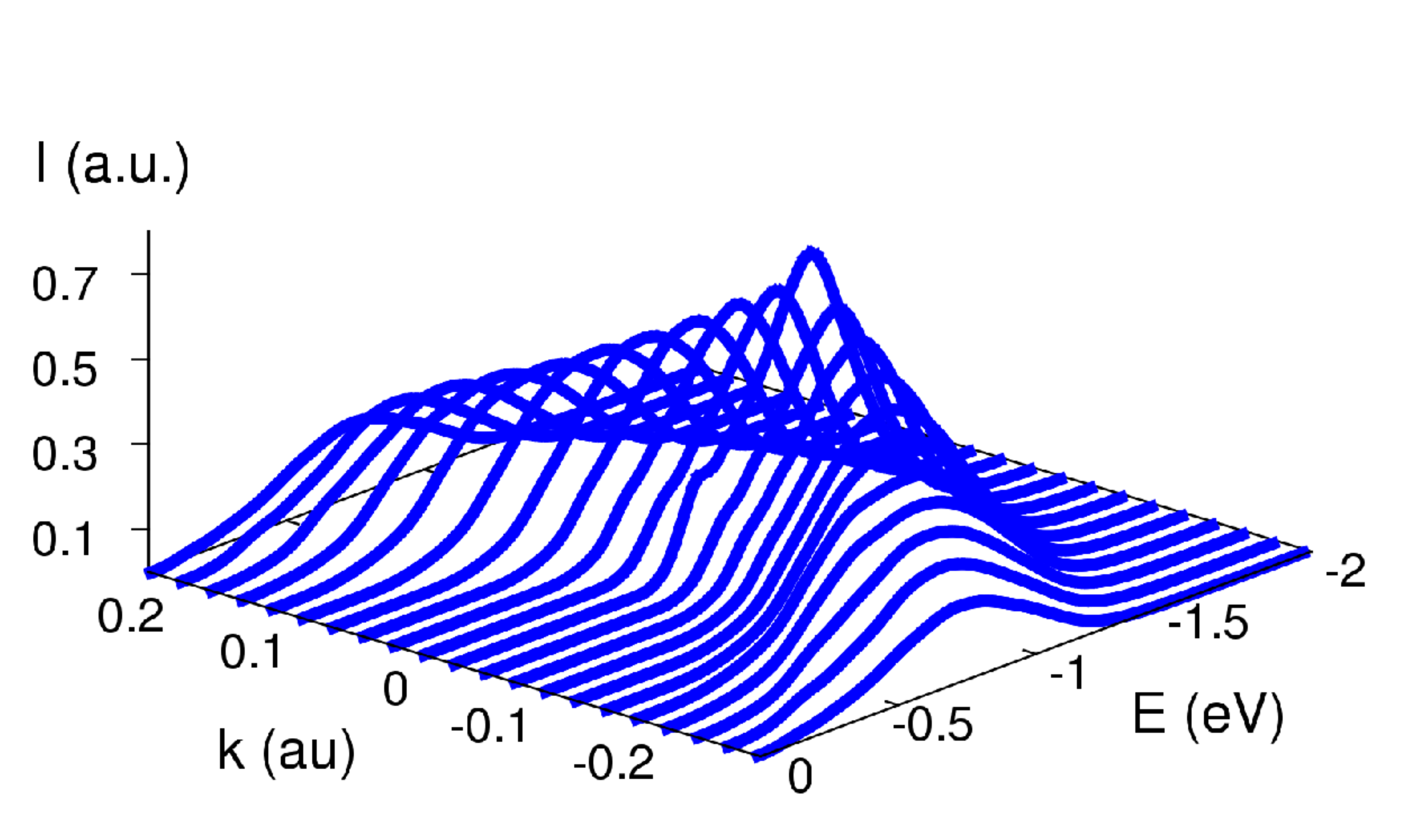}
\caption{(Color online) Photoelectron intensity as a function of $k_{\parallel}$ and E = - $\epsilon_{b}$ - $\epsilon$, where $\epsilon$ is the plasmaron excitation energy. The photon energy is 10 eV.}
\label{fig_I_k_energy}
\end{figure}

For the system p(2$\times$ 2)-K/Graphite, we have $k_{F}$ = 0.23 a.u., $q_{max}$=0.1 a.u., $\epsilon_{b}$ = 0.76 eV and the QW band dispersion approximately parabolic $\hbar\omega(k_{\|})$ = $\epsilon_{b}(k_{\|}/k_{F})^{2}$. The calculated ARPES intensity according to Eq. (\ref{eq_Wkomega}) gives the $extrinsic$ plasmaron excitations and is shown as function of parallel momentum and energy in Fig. \ref{fig_I_k_energy}. The peak at $k_{\parallel}$=0 appears at about 1.29 eV below the Fermi level which is 0.53 eV below the bottom of the QW band (- $\epsilon_{b}$). It is seen in Fig. \ref{fig_I_k_energy} that the energy position of the QW band gives rise to a small kink.

\begin{figure}
\includegraphics[angle=0,origin=c,width=80 mm]{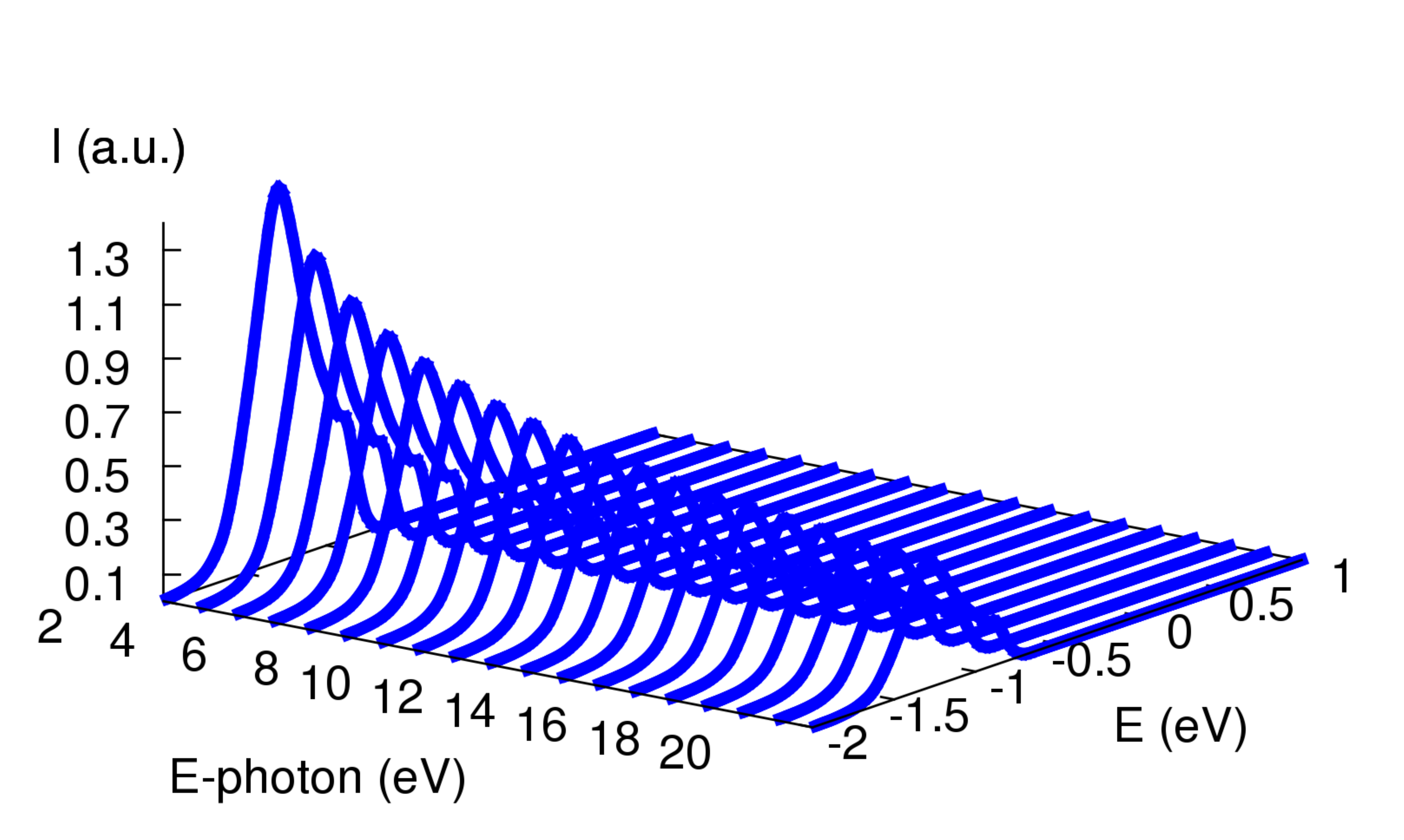}
\caption{(Color online) Photoelectron intensity in the $\bar{\Gamma}$-point ($k_{\parallel}$=0) as a function of photon energy and E = -$\epsilon_{b}$ - $\epsilon$, where $\epsilon$ is the plasmaron excitation energy.}
\label{fig_I_k_photon_energy}
\end{figure}

Calculating the photon energy dependent photo-electron yield according to Eq. (\ref{eq_I_photon}) we reveal the footprint of $extrinsic$ plasmaron excitations. The work function of p(2$\times$ 2)-K/Graphite is taken as $\phi$=2.3 eV \cite{Osterlund,Breitholtz}. In Fig. \ref{fig_I_k_photon_energy} we show the photon energy dependence of the photo-electron intensity for normal emission ($k_{\parallel}$=0).

We then fit the maximum intensity in normal emission versus photon energy $h\nu$ to a functional form given by

\begin{eqnarray}
I_{max}(h\nu) &=& A (h\nu)^{-\alpha}
\label{eq_fitt}
\end{eqnarray}

\begin{figure}
\includegraphics[angle=0,origin=c,width=80 mm]{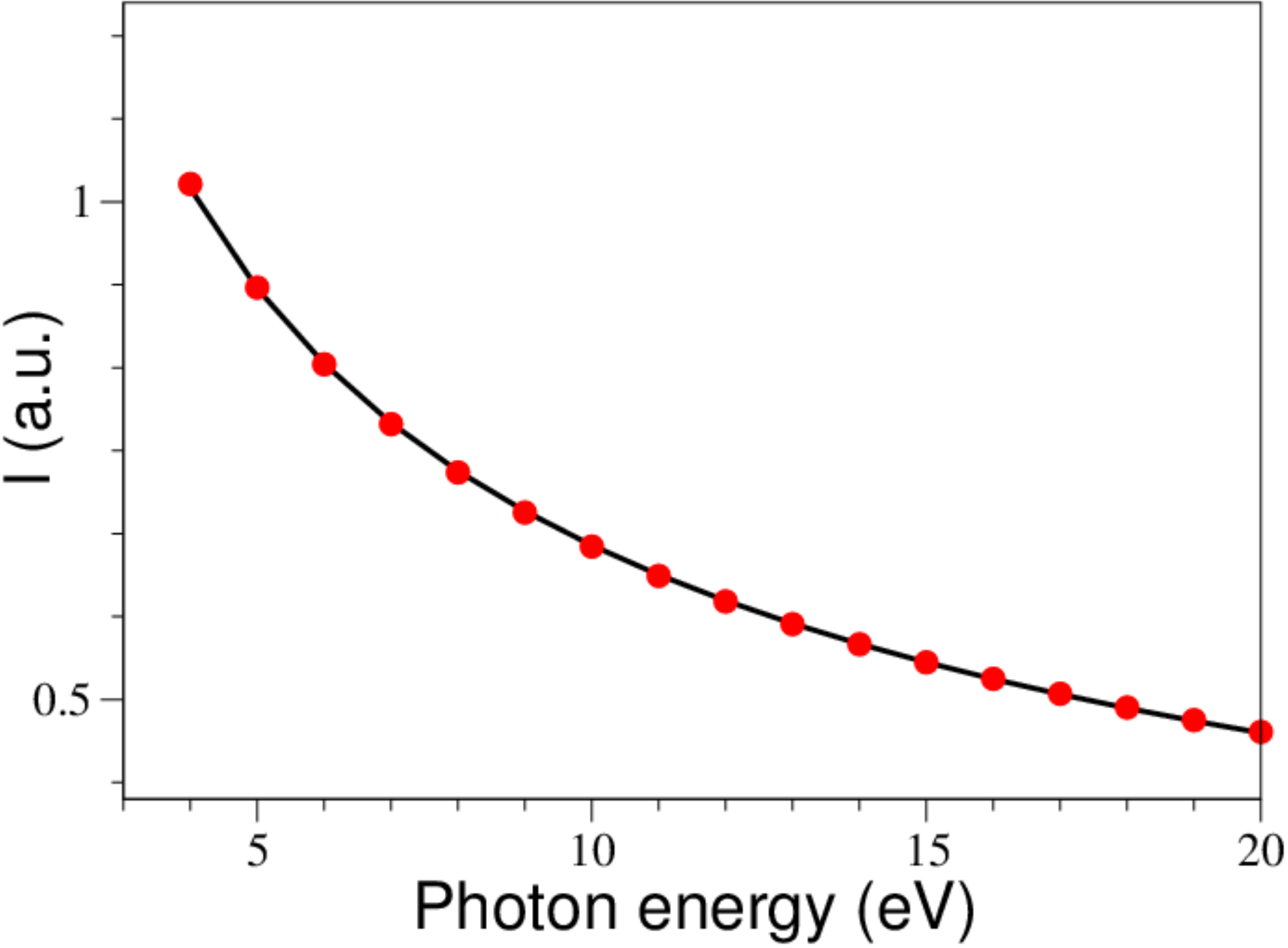}
\caption{(Color online) Solid line: Calculated photelectron yield in versus photon energy at the plasmaron peak for normal emission ($k_{\parallel}$=0). The peak is located 1.29 eV below the Fermi energy. Filled circles: Fitted  intensity versus photon energy $h\nu$ with to functional form given in Eq. (\ref{eq_fitt}) with $A$=1.99 and $\alpha$=0.49. }
\label{fig_fitted}
\end{figure}

With this fitting procedure, illustrated in Fig. \ref{fig_fitted}, we obtain $A$=1.99 a.u. and  $\alpha$=0.49. In order to have positive kinetic energy we require a minimum photon energy given by \( h\nu_{min}=\phi+\epsilon_{b}+\epsilon\) $\approx$ 3.6 eV. Thus we confirm the expected inverse square root dependence of the photo-electron intensity with respect to the photon energy as discussed previously.

\section{Summary and conclusions}

For a surface system in which excitation of acoustic surface plasmons is possible the $extrinsic$ type of plasmaron excitations are likely to exist. Referring to the photoemission experiment, the predicted footprint of the $extrinsic$ plasmarons, generated by the escaping photo-electron, is the inverse square root dependence of the photon energy. This enables a possibility to distinguish $extrinsic$ from $intrinsic$ plasmarons, where the latter is generated by the photo-hole. Following up a previous theoretical study of the system p(2$\times$2)-K/Graphite \cite{Vasse} shows that this prediction seems reliable.

\section*{ACKNOWLEDGMENTS}

V. M. S. acknowledges the partial support from the Basque Departamento de Educaci\'on, UPV/EHU (Grant No. IT-756-13) and the Spanish Ministry of Economy and Competitiveness MINECO (Grant No. FIS2013-48286-C2-1-P).

\end{document}